\documentclass[conference,a4paper]{APSIPA2019}
\usepackage{multirow}
\usepackage{graphicx}
\usepackage{amsmath}
\usepackage{subfigure}
\usepackage[psamsfonts]{amssymb}
\usepackage{amsxtra}
\usepackage{threeparttable}

\begin{document}

\title{VAE-based Domain Adaptation for Speaker Verification}

\author{%
\authorblockN{%
Xueyi Wang\authorrefmark{1}\authorrefmark{2}, Lantian Li\authorrefmark{1}, Dong Wang\authorrefmark{1}
}
\authorblockA{%
\authorrefmark{1}
Center for Speech and Language Technologies, Tsinghua University, Beijing, China\\
\authorrefmark{2}
China University of Mining \& Technology, Beijing, China
}
Corresponding Author E-mail: wangdong99@mails.tsinghua.edu.cn
}

\maketitle
\thispagestyle{empty}

\begin{abstract}

Deep speaker embedding has achieved satisfactory performance in speaker verification.
By enforcing the neural model to discriminate the speakers in the training set,
deep speaker embedding (called `x-vectors') can be derived from the hidden layers.
Despite its good performance, the present embedding model is highly domain sensitive,
which means that it often works well in domains whose acoustic condition matches that of the training data (in-domain),
but degrades in mismatched domains (out-of-domain).
In this paper, we present a domain adaptation approach based on Variational Auto-Encoder (VAE).
This model transforms x-vectors to a regularized latent space; within this latent space, a
small amount of data from the target domain is sufficient to accomplish the adaptation.
Our experiments demonstrated that by this VAE-adaptation approach, speaker embeddings can be
easily transformed to the target domain, leading to noticeable performance improvement.

\end{abstract}

\section{Introduction}

Automatic speaker verification (ASV) is an important biometric authentication technology and has found a broad range of applications.
Conventional ASV methods are based on statistical models~\cite{Reynolds00,Kenny07,dehak2011front}.
Perhaps the most famous statistical model is the Gaussian mixture model$-$universal background model (GMM-UBM)~\cite{Reynolds00}.
It factorizes the speech signal into the phonetic factor and the speaker factor, and this factorization process is based on the
\emph{maximum likelihood} (ML) criterion.
This basic factorization model was later extended to various low-rank variants, including
the joint factor analysis model~\cite{Kenny07} and the i-vector model~\cite{dehak2011front}.
Further improvements were obtained by either discriminative models (e.g., PLDA~\cite{Ioffe06}) or phonetic knowledge
transferring (e.g., DNN-based i-vector model~\cite{Kenny14,lei2014novel}).

Recently, inspired by the success of deep learning in automatic speech recognition (ASR),
the neural-based ASV models have been studied and shown great potential~\cite{ehsan14,heigold2016end,li2017deep}.
These models leverage the power of deep neural networks (DNNs) in learning strong speaker-related discriminative features,
ideally from a large amount of speaker-labelled data.
A state-of-the-art neural-based architecture is the x-vector model proposed by Snyder et al.~\cite{snyder2018xvector}.
By this architecture, frame-level deep features are derived by several full-connection layers (or more structured layers),
and then the first- and second-order statistics of frame-level features are collected and then projected to a low-dimensional representation,
which is called `x-vector'. During the training, the objective of discriminating the speakers in the training dataset encourages
the DNN structure to learn discriminative representations at both the frame level (deep feature) and the utterance level (x-vector).
The x-vector model has achieved the state-of-the-art performance in various speaker recognition tasks,
as well as related tasks such as language identification~\cite{Snyder2018}.

In spite of its powerful discriminability, the x-vector model still heavily relies on a strong back-end scoring component,
such as LDA, PLDA~\cite{Cai2018,zhang2019vae}. This is puzzling at the first glance as in the i-vector regime the back-end models play the role of
enhancing the discrimination among speaker, though the x-vectors have been discriminative already.
Our previous study shows that the back-end models play a different role when accompanying x-vectors:
instead of promoting discrimination, they essentially normalize the prior distribution of
speaker x-vectors and the conditional distribution of utterance x-vectors
of a particular speaker~\cite{zhang2019vae}.

A critical problem that usually arises in real-life applications is that the back-end models
are highly domain-sensitive, which means that an LDA-PLDA model that is well trained in one domain
may degrade significantly in other domains whose acoustic condition is
substantially different from that of the training data.
To tackle this problem, this paper presents a domain adaptation approach
based on the Variational Auto-Encoder (VAE). VAE is a powerful architecture that can
project an unconstrained distribution to a simple Gaussian distribution, and the projection
can be learned  in a purely unsupervised way. In our previous study~\cite{zhang2019vae},
VAE has been used as a normalization model that normalizes the distribution of x-vectors into a
more regularized Gaussian. This normalization, when combined with PLDA, clearly improves the ASV performance.
In this study, we investigate a
domain adaptation approach based on the VAE-based normalization architecture.
Our experiments showed that this VAE-based adaptation outperforms both the LDA- and PCA-based
adaptation and the famous unsupervised PLDA adaptation~\cite{garcia2014improving,garcia2014unsupervised}.

The organization of this paper is as follows. Section 2 describes the related work, and Section 3 presents the proposed
VAE-based adaption approach. Experiments are reported in Section 4, and the paper is concluded in Section 5.

\begin{figure*}[htb]
    \centering
    \includegraphics[width=0.95\linewidth]{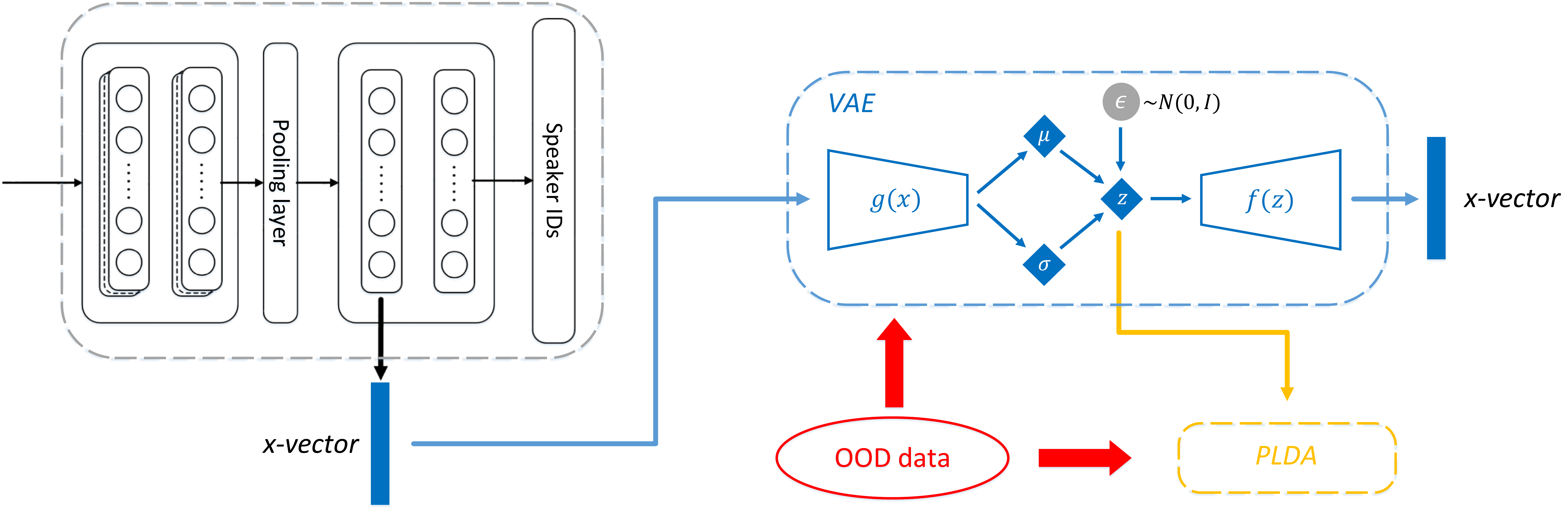}
    \caption{The three-component architecture of an x-vector system, where the normalization model is a VAE.
    X-vectors are extracted from the speaker-discriminative
    network, and then pass the VAE network for normalization. The normalized x-vectors are retrieved from the bottleneck layer of
    the VAE and scored by PLDA. The adaptation can be conducted on either VAE or PLDA, or both. }
    \label{fig:adapt}
\end{figure*}

\section{Related work}

This work is a direct extension of our previous work~\cite{zhang2019vae}.
The main contribution of this extension is that we provide a thorough investigation on the VAE-based
domain adaption for ASV.
Some recent studies on domain adaptation in the x-vector model regime are related to this work.
For example, Alam et al.~\cite{alam2018speaker} presented an unsupervised adaptation approach
based on Correlation Alignment (CORAL)~\cite{sun2016return}. CORAL can align the distributions
of in-domain and out-of-domain features. Alam et al. found that this technique can be applied to
compensate for domain mismatch of x-vectors. Lee et al. ~\cite{lee2019coral} proposed a similar
approach that employed CORAL to align the statistics of in-domain and
out-of-domain vectors. The OOD statistics was then used to update the PLDA model.
Our VAE-based approach works on the normalization model rather than the scoring model.

\section{VAE-based domain adaptation}

\subsection{Revisit VAE}
\label{sec:vae}

VAE is a generative model that can represent a complex data distribution~\cite{kingma2013auto}.
The key idea of VAE is to learn a DNN-based mapping function
$x=f(z)$ that maps a simple distribution $p(z)$ to a complex distribution $p(x)$. In other words, it represents
complex observations by simple-distributed latent codes via a complex mapping function.

In brief, VAE consists of two parts, a \textbf{decoder} $f(z)$ that maps $p(z)$ to $p(x)$, i.e.,

\[
p(x) = \int p(x|z) p(z) {\rm d}z  = \int N(f(z), I) p(z) {\rm d}z,
\]

\noindent where $p(x|z)$ has been assumed to be Gaussian. And an \textbf{encoder} $g(x)$ produces a distribution $q(z|x)$
that approximates the posterior distribution $p(z|x)$ as follows:

\[
p(z|x) \approx q(z|x) = N(\mu(x),\sigma(x)),
\]

\noindent where $[\mu(x)\ \ \sigma(x)] = g(x)$.

The training objective is the log probability of the training data $\sum_i \ln p(x_i)$.
It is intractable so a variational lower bound is optimized instead, which depends on both the encoder $g(x)$ and the decoder $f(z)$. This is formally written by:

\[
L(f,g) = \sum_i \{ -D_{\rm KL} [q(z|x_i) || p(z)] + \mathbb{E}_{q(z|x_i)} [\ln p(x_i|z)] \},
\]

\noindent where $D_{\rm KL}$ is the KL divergence, and $\mathbb{E}_{q}$ denotes expectation w.r.t. distribution $q$. As
the expectation is intractable, a sampling scheme is often used, as shown in the blue box in Fig.~\ref{fig:adapt}.
More details of the training process can be found in~\cite{kingma2013auto}.

\subsection{VAE for normalization and adaptation}

A conventional x-vector system consists of two components, one is the front-end model which is used to extract speaker embeddings (x-vectors), the other one is the back-end model which is used for scoring.
For the front-end model, it is trained by discriminating the speakers in the training set, as shown in the dotted gray box in Fig.~\ref{fig:adapt}.
To learn sufficiently discriminative and generalizable speaker embeddings, it requires a large amount of speaker-labelled data.
In spite of its powerful discriminability, the x-vector model still heavily relies on a PLDA back-end.


A potential problem, however, is that these back-end models may be domain-sensitive.
For instance, a well-trained PLDA tends to be ineffective on the out-of-domain (OOD) data.
To deal with this OOD issue, an intuitive idea is to retrain an OOD PLDA model.
However, training a PLDA model from scratch requires a large amount of labelled data,
usually thousands of speakers, each with multiple sessions. In many practical situations,
collecting such a large amount of labelled data is very difficult and time consuming.
Therefore, how to make full use of the limited speaker-labelled data is the key point
to deal with the OOD issue. To tackle this problem, a multitude of PLDA adaptation
approaches have been proposed~\cite{garcia2014improving,garcia2014unsupervised}.
In all these methods, within-class and between-class statistics are collected from
the adaptation data, and then are used to update the PLDA model.

In a previous study~\cite{zhang2019vae}, we presented a three-component architecture,
where a normalization model is introduced between the front-end (x-vector DNN) and
the back-end (PLDA). The role of the normalization model is to project x-vectors to
a latent space in which the projected codes are more regularized, e.g., more Gaussian.
This model could be PCA or LDA, but we found VAE is more powerful, due to its capability of
representing complex distributions with a simple distribution.
This three-component architecture is shown in Fig.~\ref{fig:adapt}.

This architecture motivates a new domain-adaptation approach, i.e., adapting the
normalization model rather than the PLDA back-end. In particular, there are several
advantages if we adapt the VAE-based normalization model: (1) VAE is essentially a
distribution mapping function that involves strong structural constraints (i.e., conditional Gaussian)
in both the data space and latent space. This highly structured architecture allows
effective adaptation even with a very limited amount of data;
(2) VAE training is purely unsupervised and the adaptation data are easy to obtain;
(3) After VAE adaptation, the normalized x-vectors (latent codes) remain regularized
although the distribution of raw x-vectors may have greatly changed.
This largely alleviates the necessity of PLDA adaptation. Fig.~\ref{fig:adapt} illustrates
the VAE-based adaptation, where the parameters of both VAE and PLDA could be adapted
using the OOD data.


\section{Experiments}

\subsection{Data}

Three datasets were used in our experiments: VoxCeleb, SITW and CSLT-SITW.
VoxCeleb was used for model training, while the other two were used for evaluation.
More information about these three datasets is presented below.

\emph{VoxCeleb}: A large-scale free speaker database collected by University of Oxford, UK~\cite{nagrani2017voxceleb}.
Data augmentation was applied, where the MUSAN corpus~\cite{musan2015} was used to generate noisy utterances and
the room impulse responses (RIRS) corpus~\cite{ko2017study} was used to generate reverberant utterances.
This dataset, after removing the utterances shared by SITW, was used to train the DNN x-vector model,
plus the PLDA and VAE models.

\emph{SITW-Eval.Core}: A standard free database collected by ~\cite{mclaren2016speakers} for ASV evaluation.
It was collected from open-source media channels, and consists of speech data covering $299$ well-known persons.
This dataset was used as the \textbf{IND test set}.

\emph{CSLT-SITW}: A small database collected by CSLT for commercial usage. It consists of $77$ speakers,
each of which records a simple Chinese command word, and the duration is about $2$ seconds.
The scenarios involve laboratory, corridor, street, restaurant, bus, subway, mall, home, etc.
Speakers varied their recording devices and poses during the recording.
In our experiments, $40$ speakers were used for OOD adaptation (\textbf{OOD adaptation set}),
and the rest $33$ speakers were used for OOD evaluation (\textbf{OOD test set}).

\subsection{Settings}

We built several systems to validate the VAE-based domain adaptation. All these systems use
the same x-vector front-end and PLDA back-end, but differ in the normalization model. We
denote these systems as follows.

\textbf{Baseline}: The baseline x-vector system. It was built following the Kaldi SITW recipe~\cite{povey2011kaldi}.
The feature-learning component is a $5$-layer time-delay neural network (TDNN).
The statistic pooling layer computes the mean and standard deviation of the frame-level features from a speech segment.
The size of the output layer is $7,185$, corresponding to the number of speakers in the training set.
Once trained, the $512$-dimensional activations of the penultimate hidden layer are read out as an x-vector.
There is no normalization model.

\textbf{PCA}: As the baseline, but with PCA as the normalization model. The dimension of the code space is $150$.
Similar to VAE, PCA is also an unsupervised model, though it is linear and shallow.

\textbf{LDA}: As the baseline, but with LDA as the normalization model. The dimension of the code space  is $150$.

\textbf{VAE}: As the baseline, but with VAE as the normalization model. The VAE model is a $7$-layer DNN. The dimension of code space is $200$, and other hidden layers are $1,800$.

\textbf{C-VAE}: As the baseline, but with C-VAE as the normalization model. C-VAE is a variant of VAE, with a cohesive loss involved to
encourage within-class coherence~\cite{zhang2019vae}.

\subsection{Basic results}
\label{sec:basic}

We first present the basic results evaluated on the IND test set and the OOD test set. All these three components
of the system (front-end, normalization, back-end) are trained with VoxCeleb.
The results in terms of equal error rate (EER) are reported in Table~\ref{tab:basic}.
As expected, it can be observed that for all these five systems, the performance on the IND data outperforms that on the OOD data.
For the baseline system (without any normalization), the performance
degradation on the OOD data is not much, suggesting that the DNN x-vector model has been well trained and fairly generalizable.
For systems with normalization models (PCA, LDA, VAE and C-VAE),  the performance on the IND data is significantly improved,
which confirms the contribution of normalization. However, the performance on the OOD data nearly remains unchanged,
indicating that all these normalization models suffer from a domain mismatch. In particular, the two VAE systems
drop the most on the OOD data, though their performance on the IND data is the best.
This is not surprising, as VAE/C-VAE are the most complex models and so tend to be domain-overfitting.

\vspace{-1mm}
\begin{table}[htbp]
 \begin{center}
  \caption{EER(\%) results of various systems on the IND data and the OOD data.}
   \label{tab:basic}
    \begin{tabular}{|l|c|c|c|c|c|c}
     \multicolumn{6}{c}{} \\
      \hline
                         & Baseline  & PCA  & LDA  & VAE  & C-VAE  \\
       \hline
             IND         & 16.79     &  4.84     &  3.80     &  3.64     &  3.77     \\
       \hline
             OOD         & 18.51     &  14.58    &  14.82    &  16.72    &  15.58    \\
       \hline
     \end{tabular}
 \end{center}
\end{table}
\vspace{-1mm}

\subsection{PLDA adaptation}

In this experiment, we keep all these settings as in Section~\ref{sec:basic}, but adapt the PLDA model using the
OOD adaptation data. This back-end adaptation will partly mitigate the domain-mismatch problem and so presumably
improves performance of all these systems.
We investigated two adaptation schemes: \textbf{PLDA-RET} that retrains the PLDA model from scratch,
and  \textbf{PLDA-UAT} that adapts the PLDA model using the unsupervised adaptation approach proposed by~\cite{garcia2014unsupervised}.
The results are presented in Table~\ref{tab:plda}.

\begin{table}[htb!]
 \begin{center}
  \caption{EER(\%) result on the OOD set with PLDA adaptation.}
  \vspace{-1mm}
   \label{tab:plda}
    \begin{tabular}{|l|c|c|c|c|c|c}
     \multicolumn{6}{c}{} \\
      \hline
                            & Baseline  & PCA  & LDA  & VAE  & C-VAE  \\
       \hline
             PLDA          & 18.51     & 14.58     &  14.82    &  16.72    &  15.58    \\
       \hline
             PLDA-RET      & 15.25     & 13.83     &  14.18    &  13.85    &  13.47    \\
       \hline
             PLDA-UAT      & 14.49     & \textbf{12.82} & 13.40 &  15.02    &  13.88    \\
       \hline
     \end{tabular}
 \end{center}
\end{table}
\vspace{-1mm}

Firstly, we observe that both the two PLDA adaptation approaches improve the performance on all these five systems,
as expected. Secondly, the best performance is obtained by the PCA system with PLDA-UAT.
The VAE and C-VAE systems do not work as well as the PCA and LDA systems.
This indicates that PLDA adaptation can not fully compensate for the domain-mismatch
inherence in the VAE/C-VAE models.

\subsection{Adaptation for normalization}

We have found that the normalization models, in particular VAE and C-VAE, suffer from domain-mismatch
on the OOD data, for which PLDA adaptation can not fully address.
In this experiment, we adapt both the normalization model and the PLDA back-end.
For simplicity, both these adaptations are implemented as re-training.
The results are shown in Table~\ref{tab:vae}, where `Norm-Adapt' denotes the normalization model adaptation.


\begin{table}[htbp]
 \begin{center}
  \caption{EER(\%) result on the OOD set with adaptation on both normalization models and PLDA back-end.}
   \vspace{-1mm}
   \label{tab:vae}
    \begin{tabular}{|l|c|c|c|c|c|}
     \multicolumn{5}{c}{} \\
      \hline
                                       & PCA  & LDA  & VAE  & C-VAE  \\
       \hline
           PLDA-RET                   &  13.83     &  14.18    &  13.85    &  13.47    \\
           Norm-Adapt + PLDA-RET      &  13.31    &  14.84    & \textbf{12.79} & \textbf{12.73} \\

       \hline
     \end{tabular}
 \end{center}
\end{table}
\vspace{-1mm}

Firstly, it can be observed that the adaptation on normalization models delivers
performance gains on all these systems, compared with the sole PLDA adaptation (PLDA-RET).
As expected, the improvement on the VAE and C-VAE systems is much more significant
than that on the PCA and LDA systems, indicating that adaptation
is more important for complex normalization models.
Overall, the C-VAE system obtains the best performance with both normalization model
adaptation and PLDA adaptation. This performance is better than the best unsupervised
PLDA adaptation shown in Table~\ref{tab:plda}.


\subsection{Analysis}

To better understand these adaption methods, we compute the skewness and kurtosis of the distributions
of normalized x-vectors of utterances in the OOD test dataset. The skewness and kurtosis are defined as follows:

\[
{\rm Skew}(x) = \frac {E[(x-\mu_x)^3]}{\sigma_x^3} \ , \ {\rm Kurt}(x)=\frac{E[x-\mu_x]^4}{\sigma_x^4} - 3,
\]

\noindent where $\mu_x$ and $\sigma_x$ denote the mean and standard variation of $x$, respectively.
The more Gaussian a distribution is, the closer to zero the two values are.

The utterance-level skewness and kurtosis of x-vectors normalized by different normalization models
are reported in Table~\ref{tab:gauss}.
The \textbf{Original} group denotes the normalized vectors produced by the original normalization models trained with VoxCeleb,
and the \textbf{Adaptation} group denotes the normalized vectors produced by the adapted normalization models.

\begin{table}[htbp]
  \centering
  \caption{Utterance-level Skewness and Kurtosis of x-vectors normalized by different normalization models.}
  \vspace{-1mm}
   \label{tab:gauss}
    \begin{tabular}{|c|c|c|c|}
    \hline
                          &    & \multicolumn{1}{c|}{Skew}    & \multicolumn{1}{c|}{Kurt} \\
    \hline
    \multirow{5}{*}{\renewcommand{\multirowsetup}{\centering}Original} & Baseline & -0.0890 & -0.1154 \\
     \cline{2-4}          & PCA & 0.0004 & 0.0713   \\
     \cline{2-4}          & LDA & 0.0050 & 0.1257   \\
     \cline{2-4}          & VAE & 0.0096 & 0.0560   \\
     \cline{2-4}          & C-VAE & -0.0132 & -0.0027 \\
    \hline
    \multirow{4}{*}{Adaptation} & PCA & -0.0076 & 0.1447   \\
     \cline{2-4}          & LDA & 0.0054 & 0.3465    \\
     \cline{2-4}          & VAE & \textbf{-0.0010} & \textbf{-0.0115}  \\
     \cline{2-4}          & C-VAE & \textbf{-0.0023} & \textbf{0.0011}   \\
    \hline
    \end{tabular}%
  \label{tab:addlabel}%
\end{table}%

In the original group, the skewness and kurtosis values of the utterance-level x-vectors are clearly
reduced by any of the normalization models, confirming that PCA, LDA, VAE and C-VAE are capable of normalizing x-vectors.
Moreover, it can be found that the skewness and kurtosis of the PCA and LDA normalized vectors are
smaller than VAE and C-VAE normalized vectors. This  indicates
that in the OOD scenario, PCA and LDA can do better than VAE and C-VAE in vector normalization. This is consistent with the observations
in Table~\ref{tab:plda}, where the PCA and LDA systems perform better than the VAE and C-VAE systems on the OOD data.

After adaptation, the skewness and kurtosis of VAE and C-VAE normalized vectors are clearly reduced.
This is understandable as these two models are the most powerful in
distribution normalization. This normalization does not work well on the OOD data, but a simple adaptation will recover the power quickly.
Again, these results are consistent with the observations shown in Table~\ref{tab:vae}, where VAE and C-VAE show the best
performance after adaptation.

\section{Conclusions}

This paper proposed a VAE-based domain adaptation approach for deep speaker embedding.
VAE (and its variant C-VAE) is a powerful model for normalizing the distribution of x-vectors,
and can be easily adapted to a new domain with a small amount of data.
Experiments demonstrated that this VAE-based adaptation outperforms the
LDA- and PCA-based adaptation, and when combined with PLDA re-training, it outperforms
the unsupervised PLDA adaptation.

\section*{Acknowledgement}

This work was supported by the National Natural Science Foundation of China No. 61633013, and the Postdoctoral Science Foundation of China No. 2018M640133.

\bibliographystyle{IEEEbib}
\bibliography{refs}

\end{document}